\documentclass[12pt]{article}

\usepackage{times}
\usepackage{graphicx}

\begin{document}

\baselineskip   24pt
\topmargin      0.0cm
\oddsidemargin  0.2cm
\textwidth      16cm 
\textheight     21cm
\footskip       1.0cm

\newcommand{\tagin}{\mathrm{in}}
\newcommand{\tagout}{\mathrm{out}}
\newcommand{\tagtele}{\mathrm{tele}}
\newcommand{\tagcat}{\mathrm{cat}}

\newcommand{\ketphi}{|\phi\rangle}
\newcommand{\ketpsi}{|\psi\rangle}
\newcommand{\ketin}{|\phi^\tagin\rangle}
\newcommand{\brain}{\langle\phi^\tagin|}
\newcommand{\ketvac}{|0\rangle}
\newcommand{\palpha}{|\alpha\rangle} 
\newcommand{\malpha}{|-\alpha\rangle} 
\newcommand{\ketcat}{|\tagcat\rangle}
\newcommand{\bracat}{\langle\tagcat|}

\newcommand{\ket}[1]{|#1\rangle}
\newcommand{\bra}[1]{\langle#1|}
\newcommand{\bravac}{\bra{0}}
\newcommand{\Win}{W_{\textrm{in}}}
\newcommand{\Wout}{W_{\textrm{out}}}

\def\thesection{\arabic{section}}


\title{Teleportation of Nonclassical Wave Packets of light}

\author{
Noriyuki Lee$^{1}$, 
Hugo Benichi$^{1}$, 
Yuishi Takeno$^{1}$, 
Shuntaro Takeda$^{1}$,\\ 
James Webb$^{2}$, 
Elanor Huntington$^{2}$ 
and Akira Furusawa$^{1\ast}$\\
\normalsize{$^{1}$Department of Applied Physics, School of Engineering},
\normalsize{The University of Tokyo,}\\
\normalsize{7-3-1 Hongo, Bunkyo-ku, Tokyo 113-8656, Japan} \\
\normalsize{$^{2}$Centre for Quantum Computation and Communication Technology,}\\ 
\normalsize{School of Engineering and Information Technology,}  \\
\normalsize{University College, The University of New South Wales, Canberra ACT 2600, Australia} \\
\normalsize{$^\ast$To whom correspondence should be addressed; E-mail:  akiraf@ap.t.u-tokyo.ac.jp.}
}

\date{}

\maketitle

\begin{quote}
{\bf
We report on the experimental quantum teleportation of strongly nonclassical wave packets of light.
To perform this full quantum operation while preserving and retrieving the fragile non-classicality of the input state, we have developed a broadband, zero-dispersion teleportation apparatus that works in conjunction with time-resolved state preparation equipment.
Our approach brings within experimental reach a whole new set of hybrid protocols involving discrete- and continuous-variable techniques in quantum information processing for optical sciences.
}
\end{quote}


In the early development of quantum information processing (QIP),  a communication protocol called quantum teleportation was discovered ({\it 1}) that involves the transportation of an unknown arbitrary quantum state $\ketpsi$ by means of entanglement and classical information.
Experimental realizations of quantum teleportation ({\it 2, 3}) and more advanced related operations ({\it 4})  in the continuous-variable regime have been achieved by linear optics methods, although only for Gaussian states so far. 
However, at least third order nonlinear operations are necessary for building a universal quantum computer ({\it 5})---something that Gaussian operations and Gaussian states alone cannot achieve.
Photon subtraction techniques based on discrete-variable technology can provide useful nonlinearities and are used to generate Schr\"{o}dinger's-cat states and other optical non-Gaussian states ({\it 6}).
Schr\"{o}dinger's-cat states are of particular interest in this context, as they have been shown to be a useful resource for fault-tolerant QIP ({\it 7}).
It is therefore necessary to extend the continuous-variable technology to the technology used in the world of non-Gaussian states.

We have combined these two sets of technologies, and here we demonstrate such Gaussian operations on nonclassical non-Gaussian states by achieving experimental quantum teleportation of Schr\"{o}dinger's-cat state{s} of light.
Using the photon-subtraction protocol, we generate quantum states closely approximating Schr\"{o}dinger's-cat state{s} in a manner similar to ({\it 8-11}).
To accommodate the required time-resolving photon detection techniques and handle the wave-packet nature of these optical Schr\"{o}dinger's-cat states, we have developed a hybrid teleporter built with continuous-wave light yet able to directly operate in the time domain.
For this purpose we constructed a time-gated source of Einstein-Podolsky-Rosen (EPR) correlations as well as a classical channel with zero phase dispersion ({\it 12}).
We were able to bring all the experimental parameters up to the quantum regime,
and we performed successful quantum teleportation in the sense that both our input and output states are strongly nonclassical.

A superposition of the quasi-classical coherent state $\palpha$ is one of the consensus definitions of a Schr\"{o}dinger's-cat state $\ketcat$, typically written $\ketcat\propto\palpha\pm\malpha$.
Such optical Schr\"{o}dinger's-cat states are known to be approximated by multiple photon subtractions from a squeezed vacuum state ({\it 6}).
In these protocols, a squeezed vacuum state $\hat{S}(s)\ketvac$ is weakly tapped via a subtraction channel, where $\ketvac$ is the vacuum state and $\hat{S}(s)$ the squeezing operator with squeezing parameter $s$.
When a photon detection event occurs in the subtraction channel, $\hat{S}(s)\ketvac$ is projected by the quantum action of the photon detector onto a non-Gaussian state, which can be tuned to approximate a Schr\"{o}dinger's-cat state ({\it 8-10}).
The approximation is not perfect and can be quantified by means of the fidelity figure $F_\tagcat = | \bracat \hat{a}\hat{S}(s)\ketvac |^2$ ({\it 13}).

To represent the superposition nature of these states, we use the Wigner formalism where for any quantum state $\ketphi$ one associates a quasi-probability distribution $W(x,p)$, where $x$ and $p$ are 
the phase-space position and momentum parameters. 
$W(x,p)$ is called the Wigner function and holds information exactly equivalent to $\ketphi$ ({\it 14}). 
Although the position $\hat{x}$ and momentum $\hat{p}$ quadratures operators and the vector state $\ketphi$ are abstract objects, $W(x,p)$ is always a definite real-valued function that can be numerically reconstructed if one performs a complete phase-resolved sequence of homodyne measurement $\hat{x}\cos{\theta} + \hat{p}\sin{\theta}$, a process called quantum tomography ({\it 15, 16}). 
$W(x,p)$ is not a true probability distribution, however, as there exist quantum states whose Wigner functions are not positive.
$\ketphi$ is defined to be a strongly nonclassical state when its Wigner function $W(x,p)$ fails to be a positive distribution.
Negativity in $W(x,p)$ turns out to be an especially useful description of the nonclassicality of a Schr\"{o}dinger's-cat state $\ketcat$;
$\palpha$ and $\malpha$ induce two ``classical" Gaussians in phase space, the superposition of which creates an oscillating interference pattern inducing negativity in $W(x,p)$.
In contrast, a statistical mixture of $\palpha$ and $\malpha$ would never show such negativity.

In a quantum teleportation process, the input $\Win$ and output $\Wout$ Wigner functions are related by the convolution (noted $\circ$)
\begin{equation}
  \Wout = \Win \circ G_{e^{-r}}
  \label{eq1}
\end{equation}
where $r$ is the EPR correlation parameter, $G_{\sigma}$ is a normalized Gaussian of standard deviation $\sigma$, and $\hbar$ (Plank's constant divided by 2$\pi$) has been set to 1 ({\it 17}).
When finite quantum entanglement $r$ is used, $\Wout$ will be a thermalized copy of $\Win$.
Only with infinite $r$ will $G_\sigma$ become a delta function so that $\Win = \Wout$.
The quality of quantum teleportation is usually evaluated according to the teleportation fidelity $F_\tagtele = \brain \hat{\rho}^\tagout\ketin$, which can be written as $F_\tagtele = 1/(1+e^{-2r})$ for Gaussian states ({\it 18}).
More importantly for our case, negative features of $\Win$ (if any) can only be teleported and retrieved in $\Wout$ when $F_\tagtele \geq 2/3$ ({\it 19}), a threshold also known as the no-cloning limit ({\it 20}). 
However, the practical lower bound on $F_\tagtele$ will be higher because of decoherence and experimental imperfection of $\Win$ ({\it 21}).
We have thus defined the success criterion of Schr\"{o}dinger's-cat-state teleportation as the successful transfer of its nonclassical features, or alternatively, successful teleportation of the Wigner function $\Win$ negativity.

Our experimental quantum teleporter and Schr\"{o}dinger's-cat-state source (Fig. 1) upgrade the experiments described in ({\it 3}) and ({\it 10}), respectively.
We use three optical parametric oscillators to generate the necessary squeezed vacua.
One is used for the Schr\"{o}dinger's-cat-state preparation; the other two are combined together on a half beam splitter whose two exit ports are the resulting pair of EPR-correlated beams.
The teleportation is conducted in three steps.
Alice first receives both the input state and her EPR beam and performs two joint quadrature measurements, obtaining results $x_0$ and $p_0$.
Bob then receives Alice's measurements $\beta = (x_0 + i p_0)/\sqrt{2}$ through the classical channels and applies the displacement operator $\hat{D}(\beta)$ on his EPR beam.
A final stage consists of a third homodyne detector for tomography at the teleporter output.
We emphasize that Alice and Bob do not assume any prior knowledge of the input state and adhere to unity-gain teleportation, so that the teleporter does not have any restriction regarding the specific family of quantum states it can faithfully teleport.

To benchmark our teleporter, we first evaluate the fidelity $F_\tagtele$ of teleportation of the vacuum state $\ketvac$, the coherent state of amplitude zero.
At quantum optical frequencies where the mean thermal photon number is virtually $0$, this is simply done by blocking the input port of the teleporter.
The teleported vacuum photocurrent is expected to have uniform Gaussian statistics with a variance $\sigma^2 = 1/2 + e^{-2r}$($\hbar = 1$) from which we can deduce teleportation fidelity (Fig. 2).
The blue traces are the shot-noise level, the noise spectrum of the input vacuum $\ketvac$.
The red traces are the classical limit of teleportation obtained by turning off the entanglement between Alice and Bob ($r = 0$). 
We measure 4.8 dB of added noise above the shot noise, in agreement with the expected teleportation fidelity of $0.5$.
When Alice and Bob share entanglement, the added noise drops to that shown by the green traces: 1.4 dB above the shot noise around 1 MHz, corresponding to a fidelity of $0.83$.
This is in agreement with the experimental figure of $-6.9$ dB that we observe in direct measurement of the EPR correlations shared between Alice and Bob.

In contrast to quantum teleportation experiments conducted to date for narrow sidebands of light ({\it 2, 3}), our setup operates over a wide frequency bandwidth, as required by the nature of our input state.
Because its generation relies on the detection of a single photon and the induced projection, a Schr\"{o}dinger's-cat state made via photon subtraction is a short wave packet of light.
A phenomenological way to picture these wave packets is to consider them as the closed boxes containing the macroscopic superposition states as in Schr\"{o}dinger's original idea.
This requires Alice and Bob to teleport every frequency component of these ``box-like" wave packets for faithful teleportation to occur.
In this way, Alice and Bob do not need to actually teleport the Schr\"{o}dinger's-cat states directly, but merely the potential boxes containing them.
Consequently, Alice and Bob do not need to know when a detection event occurs; rather, they are only concerned with continuous and faithful ``box" wave-packet teleportation, whichever state lies in the box.
In fact, Alice and Bob actually teleport most of the time a squeezed vacuum state $\hat{S}(s)\ketvac$.

In essence, our teleporter is a time-resolved apparatus that deconstructs the input wave packets into a stream of infinitely small time bins and reconstructs them at the output, within the extent of what we refer to as the teleportation bandwidth.
This bandwidth is clearly visible in both green experimental traces where the added noise slowly increases with frequency (Fig. 2).
This is a direct consequence of the finite bandwidth of squeezing used for entanglement.
However, across the frequencies relevant to our input state, teleportation fidelity is always greater than the no-cloning limit of $2/3$, a necessary regime for negativity teleportation.
A very careful implementation of the classical channel has been required ({\it 12}) to achieve experimental realization of this fidelity.

To verify the success of Schr\"{o}dinger's-cat-state teleportation, we perform experimental quantum tomography of the input and output states independently (Fig. 3).
Both input and output marginal distributions exhibit the characteristic eye shape of photon-subtracted squeezed states, with a clear lack of detection events around the origin for any phase.
Although necessary, this feature alone is not sufficient to confirm the presence of negativity in $\Win$ or $\Wout$.
The reconstructed input Wigner function $\Win$ shows the two positive Gaussians of $\palpha$ and $\malpha$ together with a central negative dip ($\Win(0,0) = -0.171\pm0.003$) caused by the interferences of the $\palpha$ and $\malpha$ superposition.
The output Wigner function $\Wout$ retains the characteristic non-Gaussian shape as well as the negative dip ($\Wout(0,0) = -0.022\pm0.003$) to a lesser degree.
The degradation of the central negative dip and the full evolution of $\Win$ towards $\Wout$ can be fully understood using Eq. 1 with a model of $\Win$, as was done in ({\it 21}).
Given the measured input state negativity of $\Win(0,0) = -0.171$ and $-6.9$ dB of squeezing,
the results of ({\it 21}) predict an output negativity value of $-0.02$, in good agreement with measured output negativity. 
Although this figure does not take into account the input-state squeezing, a more detailed model shows that a squeezing parameter $s=0.28$ affects output negativity in the third decimal place only ({\it 12}).
The experimental input and output states have an average photon number $\langle \hat{n} \rangle$ equal to $1.22\pm0.01$ and $1.33\pm0.01$, respectively ({\it 12}).
The increase in the output-state size is due to teleportation induced thermalization.
We calculate that the fidelity $F_\tagcat$ is as high as $0.750\pm0.005$ for the input Wigner function $\Win$, with the nearest Schr\"{o}dinger's-cat having an amplitude $|\alpha_\tagin|^2 = 0.98$ ({\it 12}).
However, after the teleportation $\Wout$, fidelity is reduced to $0.46\pm0.01$, with the nearest Schr\"{o}dinger's-cat state having an amplitude $|\alpha_\tagout|^2 = 0.66$. 
If $\Wout$ fidelity is calculated with $|\alpha_\tagin|^2 = 0.98$, then $F_\tagcat = 0.45\pm0.01$. 

We have demonstrated an experimental quantum teleporter able to teleport full wave packets of light up to a bandwidth of 10 MHz while at the same time preserving the quantum characteristic of strongly nonclassical superposition states, manifested in the negativity of the Wigner function.
Although $F_\tagcat$ and $W(0,0)$ drop in the teleportation process, there is no theoretical limitation other than available squeezing, and stronger EPR correlations would achieve better fidelity and negativity transmission.
The various more complex states generated as an application of photon-subtraction so far ({\it 22, 23}) can be, in principle, readily sent through our broadband quantum teleporter. 
This opens the door to universal QIP and further hybridization schemes between discrete- 
and continuous-variable techniques ({\it 24}).


\clearpage

\section*{Figure Captions}

\subsection*{Figure 1}

{\bf Experimental setup.}
OPO: Optical Parametric Oscillator,
APD: Avalanche Photo Diode,
HD: Homodyne Detector,
LO: Local Oscillator,
EOM: Electro-Optical Modulator,
ADC: Analog to Digital Converter,
FC: Filtering Cavity. 
See ({\it 12}) for details.

\subsection*{Figure 2}

{\bf Broadband teleportation of the vacuum state $\ketvac$.}
Experimentally measured power spectra of the measured photocurrents calculated by Fourier transform are shown for the position (\textbf{A}) and momentum (\textbf{B}) quadratures.
(\textbf{blue}) shotnoise input,
(\textbf{green}) quantum teleportation output,
(\textbf{red}) teleportation output without entanglement.
Reconstructed Wigner functions of the input state $\ketvac$ (\textbf{C}), quantum teleported vacuum (\textbf{D}) and classically teleported vacuum (\textbf{E}).

\subsection*{Figure 3}

{\bf Teleportation of Schr\"{o}dinger's cat states.}
Experimentally measured input state's Wigner function $\Win$ (\textbf{A}), 
marginal distribution (\textbf{B}) and photon number distribution (\textbf{C}),
as well as output state's Wigner function $\Wout$ (\textbf{D}),
marginal distribution (\textbf{E}) and photon number distribution (\textbf{F}).

\clearpage

\section*{References and Notes}

\begin{itemize}

	\item[1.]
	C. H. Bennett {\it et al.},
  {\it Phys.\ Rev.\ Lett.\/} {\bf 70}, 1895-1899  (1993),
  L. Vaidman, {\it Phys.\ Rev.\ A.} {\bf 49}, 1473 (1994).

	\item[2.]
	A. Furusawa {\it et al.},
	{\it Science} {\bf 282}, 706 (1998).

	\item[3.]
	M. Yukawa, H. Benichi, A. Furusawa,
	{\it Phys.\ Rev.\ A.} {\bf 77}, 022314 (2008).

	\item[4.]
	J. Yoshikawa {\it et al.},
	{\it Phys.\ Rev.\ A.} {\bf 76}, 060301 (2007).

	\item[5.]
	S, Lloyd, S. L. Braunstein,
	{\it Phys.\ Rev.\ Lett.\/} {\bf 82}, 1784 (1999).

	\item[6.]
	M. Dakna, T. Anhut, T. Opatrn\`{y}, L. Kn\"{o}ll, D.-G. Welsch,
	{\it Phys.\ Rev.\ A.} {\bf 55}, 3184-3194 (1997).

	\item[7.]
  A. P. Lund, T. C. Ralph, H. L. Haselgrove,
	{\it Phys.\ Rev.\ Lett.\/} {\bf 100}, 030503 (2008).

	\item[8.]
	A. Ourjoumtsev, R. Tualle-Brouri, J. Laurat, P. Grangier,
	{\it Science} {\bf 312}, 83 (2006).

	\item[9.]
	 J. S. Neergaard-Nielsen, B. Melholt Nielsen, C. Hettich, K. M{\o}lmer, E. S. Polzik,
	{\it Phys.\ Rev.\ Lett.\/} {\bf 97}, 083604 (2006).

	\item[10.]
	K. Wakui, H. Takahashi, A. Furusawa, M. Sasaki,
	{\it Opt.\ Exp.\/} {\bf 15}, 3568-3574 (2007).

	\item[11.]
  A. Ourjoumtsev, H. Jeong, R. Tualle-Brouri1, P. Grangier
	{\it Nature} {\bf 448}, 784-786 (2007).

  \item[12.]
  Materials and methods are available as supporting material on {\it Science} online.

	\item[13.]
	H. Jeong, A. P. Lund, T. C. Ralph,
	{\it Phys.\ Rev.\ A.} {\bf 72}, 013801 (2005).

  \item[14.]
  M. Hillery, R. F. O'Connel, M. O. Scully, E. P. Wigner,
  {\it Phys.\ Rep.\ } {\bf 106}, 121 (1984). 

  \item[15.]
  D. T. Smithey, M. Beck, M. G. Raymer, A. Faridani,
  {\it Phys.\ Rev.\ A.} {\bf 70}, 1244 (1993).

	\item[16.]
	A. I. Lvovsky, 
	{\it J.\/ Opt.\/ B.} {\bf 54}, S556-S559 (2004).

	\item[17.]
	S. L. Braunstein, H. J. Kimble,
	{\it Phys.\ Rev.\ Lett.\/} {\bf 80}, 869-872 (1998).

	\item[18.]
	B. Schumacher,
	{\it Phys.\ Rev.\ A.} {\bf 54}, 2614-2628(1996).

	\item[19.]
	M. Ban,
	{\it Phys.\ Rev.\ A.} {\bf 69}, 054304 (2004).

	\item[20.]
	F. Grosshans, P. Grangier, 
	{\it Phys.\ Rev.\ A.} {\bf 64}, 010301(R) (2001).

  \item[21.]
  L. Mista Jr., R. Filip, A. Furusawa,
  {\it Phys.\ Rev.\ A.} {\bf82}, 012322 (2010)

	\item[22.]
  H. Takahashi {\it et al.}, 
  \emph{PRL}, \textbf{101}, 233605(2008).

	\item[23.]
	 J. S. Neergaard-Nielsen {\it et al.},
	{\it Phys.\ Rev.\ Lett.\/} {\bf 105}, 053602 (2010).

	\item[24.]
	D. Gottesman, A. Kitaev, J. Preskill,
	{\it Phys.\ Rev.\ A.} {\bf 64}, 012310 (2001).

  \item[25.]
  This work was partly supported by the SCOPE program of the MIC of Japan, SCF, GIA, G-COE, APSA, and FIRST commissioned by the MEXT of Japan, ASCR-JSPS, and the ARC COE scheme of Australia.

\end{itemize}

\clearpage

\begin{figure}
  \begin{center}
   \includegraphics[width=110mm]{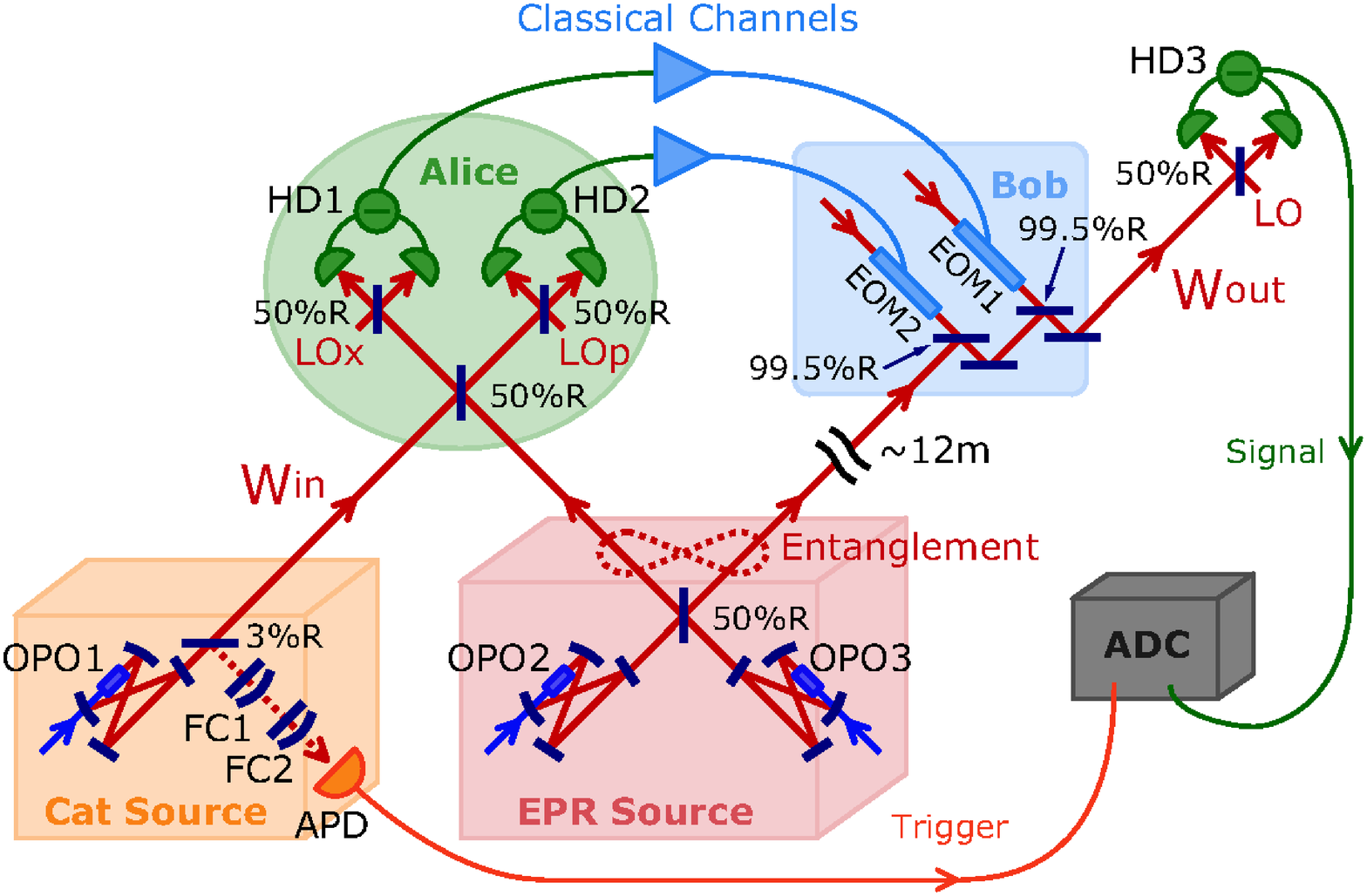}
  \end{center}
\caption{}
\end{figure}

\begin{figure}
  \begin{center}
   \includegraphics[width=110mm]{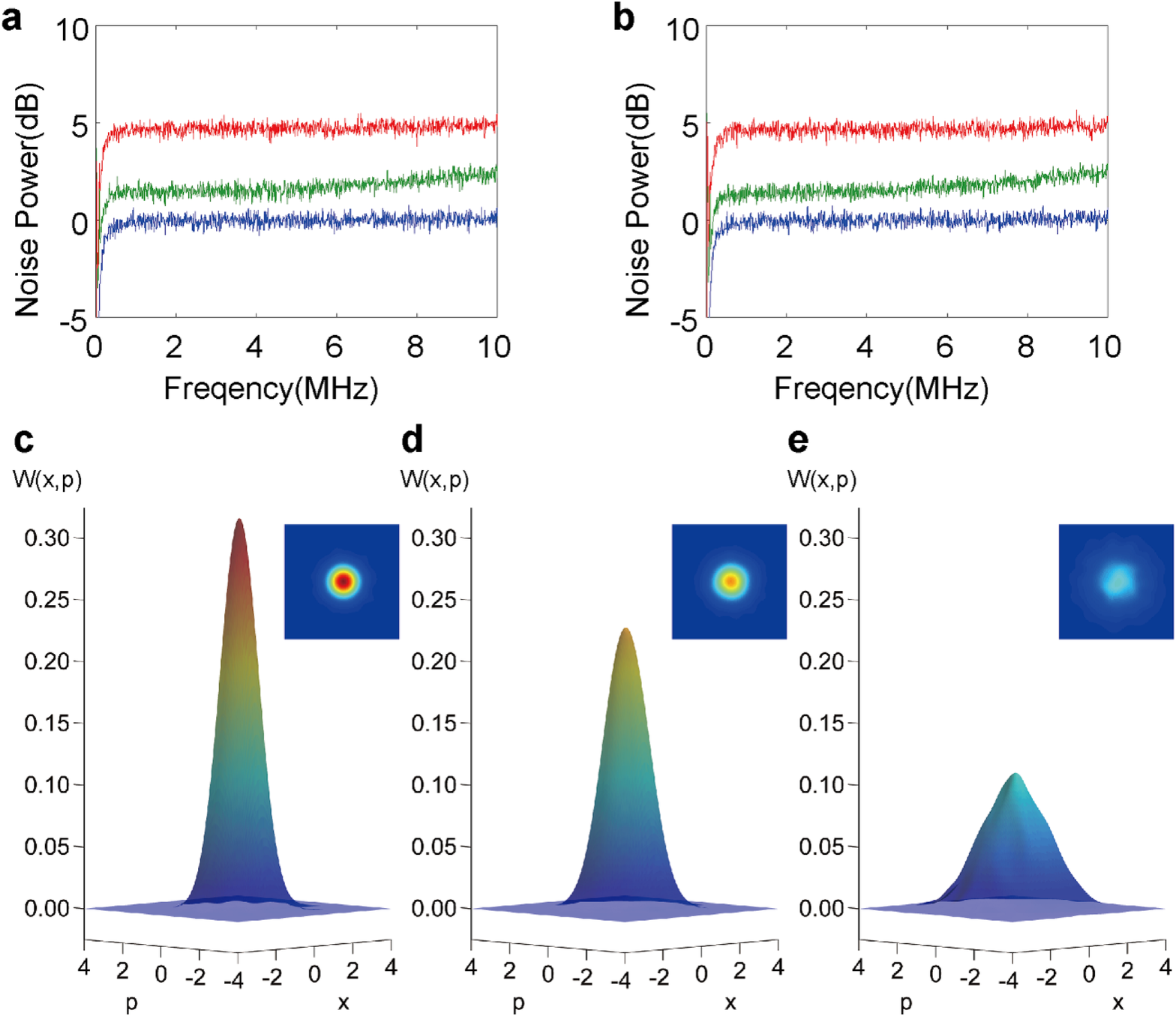}
  \end{center}
\caption{ }
\end{figure}

\begin{figure}
  \begin{center}
   \includegraphics[width=150mm]{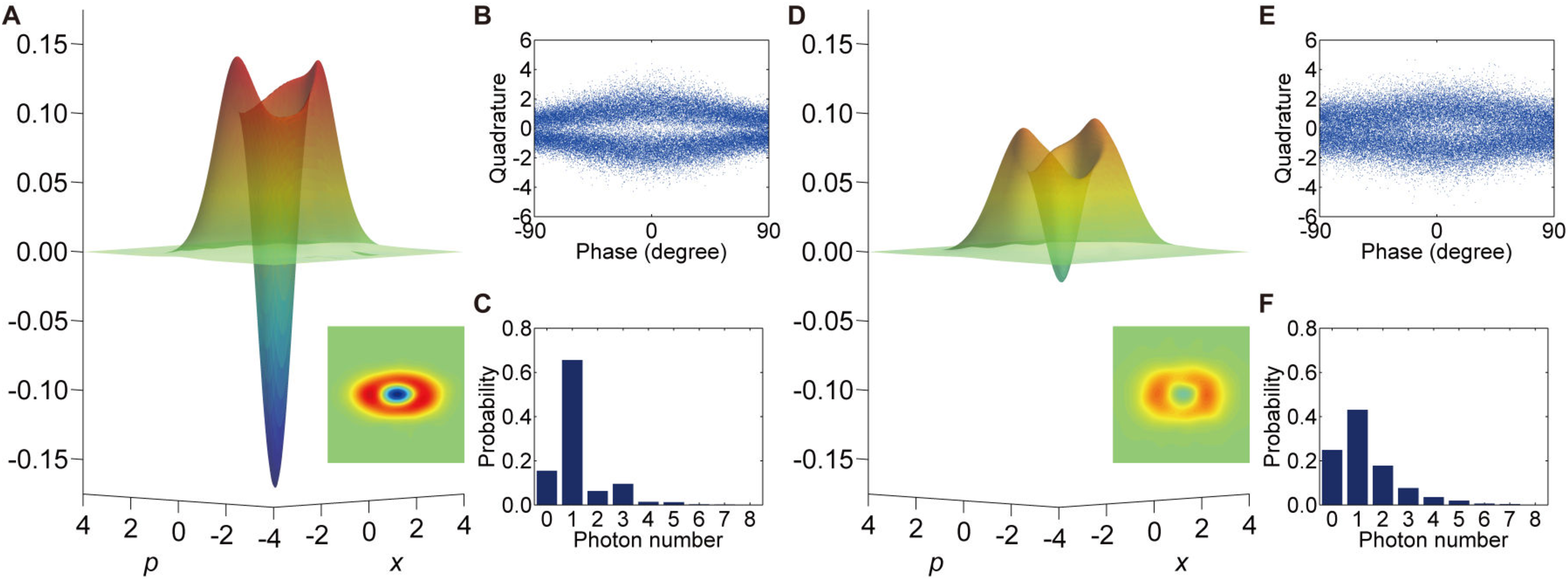}
  \end{center}
\caption{}
\end{figure}

\clearpage



\textbf{Supporting Online Material} 


\section{Experimental setup}

We use a continuous-wave titanium-sapphire laser at 860 nm (SolsTiS-SRX, M Squared Lasers, Verdi 18 W pump, Coherent) delivering a round, zero-order Gaussian TEM beam.
1 W of that beam is used to pump a second-harmonic generation cavity with a KNbO$_{3}$ crystal for squeezing generation. 
Weak coherent beams are injected into each OPO and are phase controlled with respect to each other by monitoring classical phase interference signals.
To support phase control and noiseless measurement at the same time and allow for time-gated operations, these beams are alternatively switched on and off using acousto-optical modulators at a frequency of 2 kHz.
Phase locks are achieved when the beams are switched on while measurement is performed when the beams are switched off.


\section{Schr\"{o}dinger's cat states source}

We use for the photon subtraction protocol a weakly squeezed vacuum $\hat{S}\ketvac$ generated with an Optical Parametric Oscillator (bow-tie cavity, PPKTP crystal, finesse 48, Full-Width-at-Half-Maximum 12.3 MHz).
We pump this OPO with 15 mW of 430 nm light, for a measured squeezing parameter $s = 0.28\pm0.02$. 
Intra-cavity losses with the pump injected are $0.0033\pm0.0005$.
From the produced squeezed vacuum a small fraction (3\%) is tapped on a beam-splitter and guided into an Avalanche Photo Diode (APD, Perkin-Elmer).
Two Fabry-Perot cavities (FWHM: 110 and 36 MHz, finesse: 1540 and 1630) on the way to the APD allow to select the degenerate central mode of OPO1 only.
The photon detection rate is $9000\pm500$ events per second.
The dark count rate is $120\pm20$ events per second, for a event-to-dark-counts ratio of $66\pm6$.
The actual values of $R$ and $s$ are the result of optimization between event rate and squeezing purity.
Rather than maximizing the overlap with the nearest cat state, we are actually more concerned in the input state negativity $\Win(0,0)$ since our criterion of success is negativity $\Wout(0,0)$ of the output teleported state.
By lowering the pump beam power, the purity of the squeezing is increased which is beneficial for the input state central negativity $\Win(0,0)$.
Furthermore it reduces the probability of higher order pairs of photons which is also beneficial for $\Win(0,0)$.
However it decreases the event-to-dark-counts ratio, which increases $\Win(0,0)$.


\section{Broadband teleportation}

Compared to the bandwidth of OPO1 used for the generation of Schr\"{o}dinger's cat states, we use broader bandwidth OPOs for the EPR entanglement source.
Both OPO2 and 3 are built on bow-tie cavities with PPKTP crystals.
FWHM is 24 MHz and finesse is 41.
Pump beam powers are 120 mW and intra-cavity losses with pump beam injected are respectively 0.0035 and 0.0037. 

For teleportation to work on a broadband range of frequencies, every electronic stage in the classical channel needs a flat gain band as well as a linear phase delay.
Only in that case will classical information received by Bob from Alice match in time and in magnitude the EPR correlations shared between them at every frequency. 
A nonlinear phase delay across the teleportation bandwidth would distort the wave-packet shape while a non-flat gain would result in excess noise at specific frequencies.

Bob performs displacements by mixing two auxiliary coherent beams with his EPR correlated beam on two $99.5/0.5$ beam splitters to preserve quantum correlations.
Each auxiliary beam is phase modulated according to Alice's results for the $\hat{x}$ and $\hat{p}$ quadratures.
To accommodate with the DC-10 MHz bandwidth of teleportation, we use for modulation of these auxiliary beams optical waveguide phase modulators based on Mach-Zehnder interferometers with a matched 50 $\Omega$ impedance that allows for easy broadband operations.

To compensate for the global time delay of the slower electrical classical channels, the optical path length of Bob's EPR beam is set at 12 m which allows synchronized overlap between classical information and EPR correlations owned by Bob at the output of the teleporter, both detrimental to the quality of teleportation.

Finally the gain of the classical channels is adjusted to unity following the method of ({\it 3}).


\section{Homodyne detection and tomography}

The three homodyne measurements are performed in the usual way with 50/50 beam splitters.
For better spatial mode-matching the local oscillator beams are filtered in a separate bow-tie cavity.
Visibilities are simultaneously above 99\% for every pair of signal beam and LO beam.
Homodyne detectors are built according to the transimpedance design and employ a pair of silicon pin photodiodes with a 99\% quantum efficiency (Hamamatsu, special order).
The bandwidth of the detectors is above 20 MHz. 
Local oscillator power is set at 10 mW for every homodyne measurement to avoid saturation of the photodiodes yet to obtain a high signal-to-noise ratio.

For quantum tomography, the photocurrent to be analyzed is sent to an analog to digital converter (National Instruments, PXI-5152, sampling rate 1 GHz, 8 bits resolution).
To retrieve the non-Gaussian statistics of the state the AD converter is triggered by the clicks of the APD and records 500 data points for every trigger. 
These 500 points together constitute a single quadrature measurement calculated using the temporal mode function of our input state taken as $\exp{[-\gamma |t|]}$ where $\gamma$ is the bandwidth of OPO1.
One set of measurements for a full tomography is made of 200000 quadrature measurements with a uniformly distributed phase angle $\theta$ and recorded in approximately 90 seconds.
All Wigner functions are reconstructed using the maximum-likelihood iteration algorithm without any compensation for imperfection or prior knowledge of the quantum state.


\addtocounter{equation}{1}

\section{Estimation of photon size and Schr\"{o}dinger's cat state fidelity}

From the experimentally reconstructed density matrices, we calculate the input and output state average photon number $\langle \hat{n}\rangle$ using the formula

\begin{equation}
\langle \hat{n} \rangle = {\rm tr} \left(\hat{n} \hat{\rho}\right) = \sum_{n=1}^N n \rho_{nn}
\end{equation}
where $N+1$ is the photon size of the Hilbert space used in the maximum-likelihood iteration algorithm (N = 15\footnote{For $N$ above 10 there are virtually no differences in the reconstructed states and the numerical differences have been checked to be much smaller than the experimental error boundary.}).
To calculate the Schr\"{o}dinger's cat state fidelity $F_\tagcat = \bracat \hat{\rho} \ketcat$ we express the scalar product in phase space as 
\begin{equation}
F_\tagcat = 
2\pi \int_{-\infty}^{+\infty} dx \int_{-\infty}^{+\infty} dp W_\tagcat(x,p) W^{\hat{\rho}}(x,p)
\end{equation}
where $W^{\hat{\rho}}$ is the Wigner function associated to the density matrix $\hat{\rho}$.
Our target Schr\"{o}dinger's cat state is the so-called odd-cat state $(\palpha - \malpha)/\sqrt{N}$ where $N$ is a normalization factor expressed as $2(1-\exp[-2\alpha^2])$ and $\alpha$ is a real number.
Its Wigner function $W_\tagcat$ is given as
\begin{equation}
W_\tagcat(x,p) = \frac{2}{N} e^{-x^2-p^2}
\left( e^{-2 \alpha^2} \cosh(2\sqrt{2}\alpha\,x) - \cos( 2\sqrt{2}\alpha\,p )\right)
\end{equation}
To estimate the fidelity to the nearest odd-cat state, using our reconstructed density matrix $\hat{\rho}$ and the above expression of $W_\tagcat$ we evaluate numerically the integral (3)  and look for its maximum value over the free parameter $\alpha$.


\section{Evolution of negativity in the teleportation}

Using equation (1) we want to predict the expected negativity at the output of the teleporter given the EPR correlation parameter $r$ and the input Wigner function $\Win$.
As $\Win$ is neither a pure state nor a Gaussian state there is no easy way to characterize it and we need to use a theoretical model to approximate the experimentally measured $\Win$.
The model proposed in ({\it 21}) uses a mixture of vacuum $\ketvac$ and one photon state $\ket{1}$ written
\begin{equation}
\rho_1 = \eta \ket{1}\bra{1} + (1-\eta) \ketvac \bravac
\end{equation}
If $W^{(1)}(x,p)$ is the Wigner function of $\rho_1$, then the negativity of this state is expressed by $W^{(1)}(0,0) = (1-2\eta)/\pi$. 
Using equation (1) we derive the negativity of the output state Wigner function, had $\rho_1$ been used as the input state:
\begin{equation}
W^{(1)}_{{\rm out}}(0,0) = \left( 1 - 2 \eta + 2 e^{-2r} \right) / \pi \left( 1 + 2 e^{-2r} \right)
\end{equation}
Using the measured input state negativity we calculate $\eta=0.77$. From the measured squeezing of OPO2 and 3 we deduce $r=0.795$, from which we calculate $W^{(1)}_{{\rm out}}(0,0) = -0.0207$.
We need a more advanced model to include the effect of the squeezing of the input state (quantified by the squeezing parameter $s$) on the output state negativity.
For that we first think of the input state as the state $\ket{\phi_2} = \hat{a}\hat{S}(s)\ketvac$ whose Wigner function $W^{(2)}$ is written
\begin{equation}
W^{(2)}(x,p) = 2(e^{2s}x^2+e^{-2s}p^2-1/2) \exp \left[-e^{2s}x^2 - e^{-2s}p^2 \right]
\end{equation}
However $\ket{\phi_2}$ is a pure state and does not model the experimental measured Wigner function $\Win$ well. 
Especially it has maximum negativity $-1/\pi$.
We model $\Win$ using $W^{(2)}(x,p)$ together with the linear beam splitter loss model with loss parameter $1-\eta$:
\begin{equation}
  W^{(3)}(x,p) = \frac{1}{\eta}\left(W^{(2)} \ast G_\lambda\right)\left(\frac{x}{\sqrt{\eta}},\frac{p}{\sqrt{\eta}}\right)
\end{equation}
with $\lambda = \sqrt{\frac{1-\eta}{2\eta}}$ and $G_{\sigma}$ a normalized Gaussian of standard deviation $\sigma$.
Using again equation (1) on $W^{(3)}$ we finally derive the output negativity $W^{(3)}_{{\rm out}}(0,0)$ expressed as
\begin{equation}
W^{(3)}_{{\rm out}}(0,0) 
= \frac{g_r(g_r-2\eta)}{\pi\left(g^2_r+4\eta(g_r-\eta){\rm sh}^2(s)\right)^{3/2}} 
\end{equation}
with $g_r = 1+2e^{-2r}$.
Using the same experimental figures as before and adding $s = 0.28$, we deduce $\eta = 0.79$, $r= 0.795$ and $W^{(3)}_{{\rm out}}(0,0) = -0.0247$.
We notice that the output negativity is increased when using the second model of input state.
This can be understood when looking at the higher value of $\eta$ needed to fit the input negativity $\Win(0,0)$.
Therefore $W^{(3)}$ is actually a purer state than $W^{(1)}$ and is easier to teleport in term of negativity.


\end{document}